\documentclass[a4paper]{PoS}

\title{Seyfert galaxies with \textit{Swift}: giant flares, rapid drops, and other surprises}

\ShortTitle{Seyfert galaxies with Swift}

\author{S. Komossa\\ 
Max-Planck-Institut f{\"u}r Radioastronomie, Auf dem H{\"u}gel 69, 53121 Bonn, Germany; 
E-mail: \email{skomossa@mpifr.de}}

\author{D. Grupe\\
Department of Earth and Space Science, Morehead State University, 235 Martindale
Dr., Morehead, KY 40351, USA\\} 

\author{R. Saxton\\
XMM SOC, ESAC, Apartado 78, 28691 Villanueva de la Ca{\~n}ada, Madrid, Spain\\}

\author{L. Gallo\\
Department of Astronomy and Physics, Saint Mary's University, 923 Robie Street, Halifax, NS B3H 3C3, Canada\\}

\abstract{
\textit{Swift} has initiated a new era of understanding the extremes
of active galactic nuclei (AGN) variability, their drivers and underlying physics.
This is based on its rapid response, high sensitivity,
good spatial resolution, and its ability to collect simultaneously
X--ray-to-optical SEDs.
Here, we present results from our recent monitoring campaigns with
\textit{Swift} of highly variable AGN, including outbursts, deep low states,
and unusual long-term trends in several Seyfert galaxies
including Mrk\,335, WPVS\,007, and
RXJ2314.9+2243. We also report detection of a new X-ray
and optical outburst of IC\,3599
and our \textit{Swift} follow-ups. IC\,3599 was previously known as one of the
AGN with the highest-amplitude outbursts. We briefly discuss
implications of this second outburst of IC\,3599 for emission scenarios including
accretion-disk variability, repeat tidal disruption events, and
the presence of a binary supermassive black hole.
}

\FullConference{Swift: 10 Years of Discovery\\
2-5 December 2014\\
La Sapienza University, Rome, Italy}

\begin{document}

\section{Introduction}

AGN variability provides us with a powerful tool of understanding
the physics of the central engine (e.g., Fabian 2013). While many AGN show
mild variability, a few of them exhibit high-amplitude outbursts
or low states, changing their flux by factors 20-50, or in very rare
cases exceeding factors of 100.
Spectral changes during these extreme states inform us about
the processes near the supermassive black hole (SMBH),
including reflection and relativistic effects, accretion physics,
and the presence and properties of cold and highly ionized absorbers.
In order to catch AGN in extreme states,
multi-wavelength long-term monitoring, and dedicated follow-ups once
a source flares or dips, are required. The \textit{Swift} mission
(Gehrels et al. 2004) has provided important
contributions to this field.
It has served as both: the discovery mission, triggering
deep follow-ups with other instruments; or as follow-up mission,
providing invaluable information on the lightcurve evolution of peculiar
states of sources.

\begin{figure*}[b]
\centering
\includegraphics[width=3.1in]{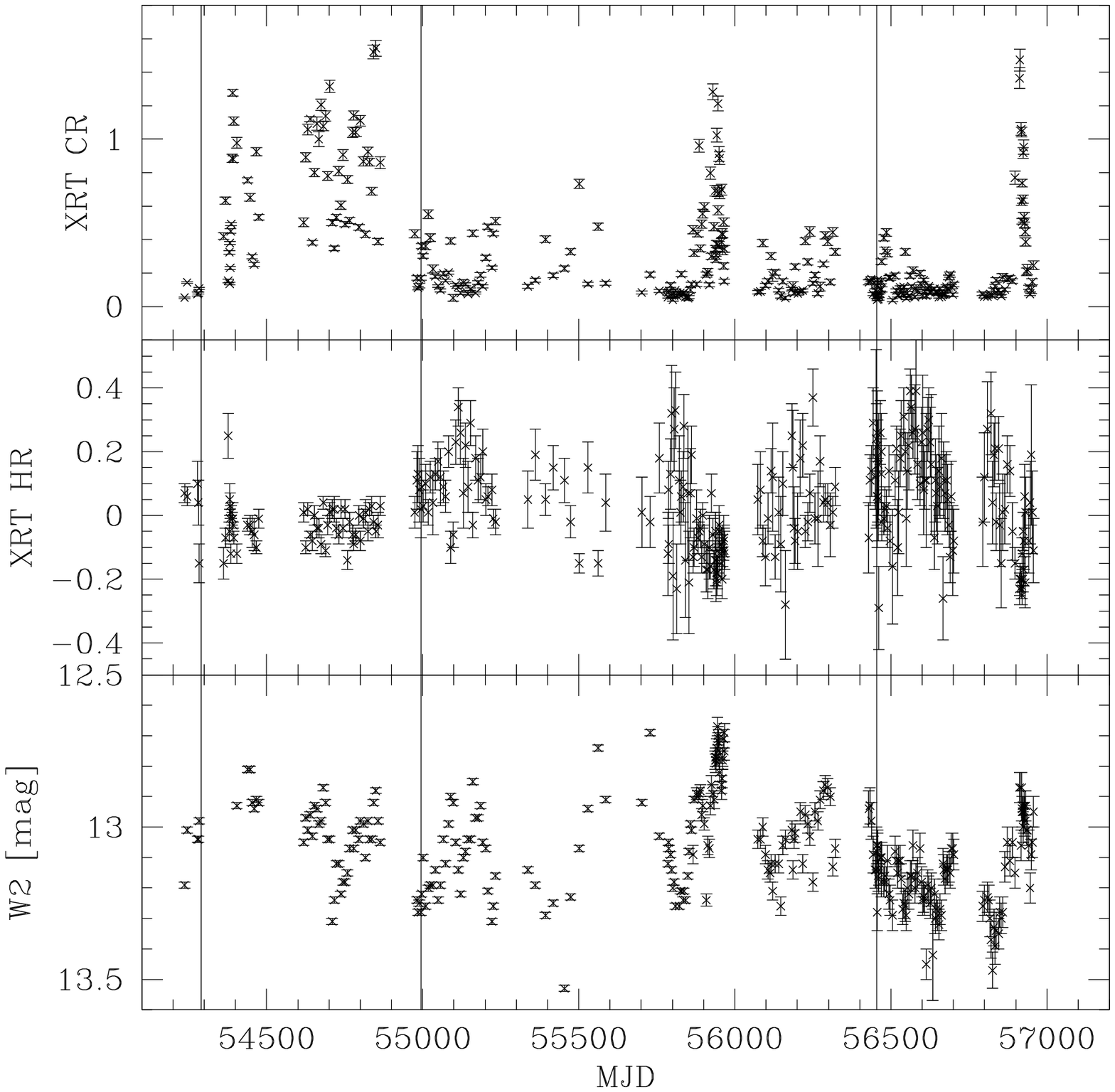}
\caption{\textit{Swift} long-term lightcurve of Mrk 335 (upper panel: X-ray telescope (XRT)
countrate, middle panel: hardness ratio,
lower panel: UVOT W2 magnitude).}
\centering
\includegraphics[width=3.0in]{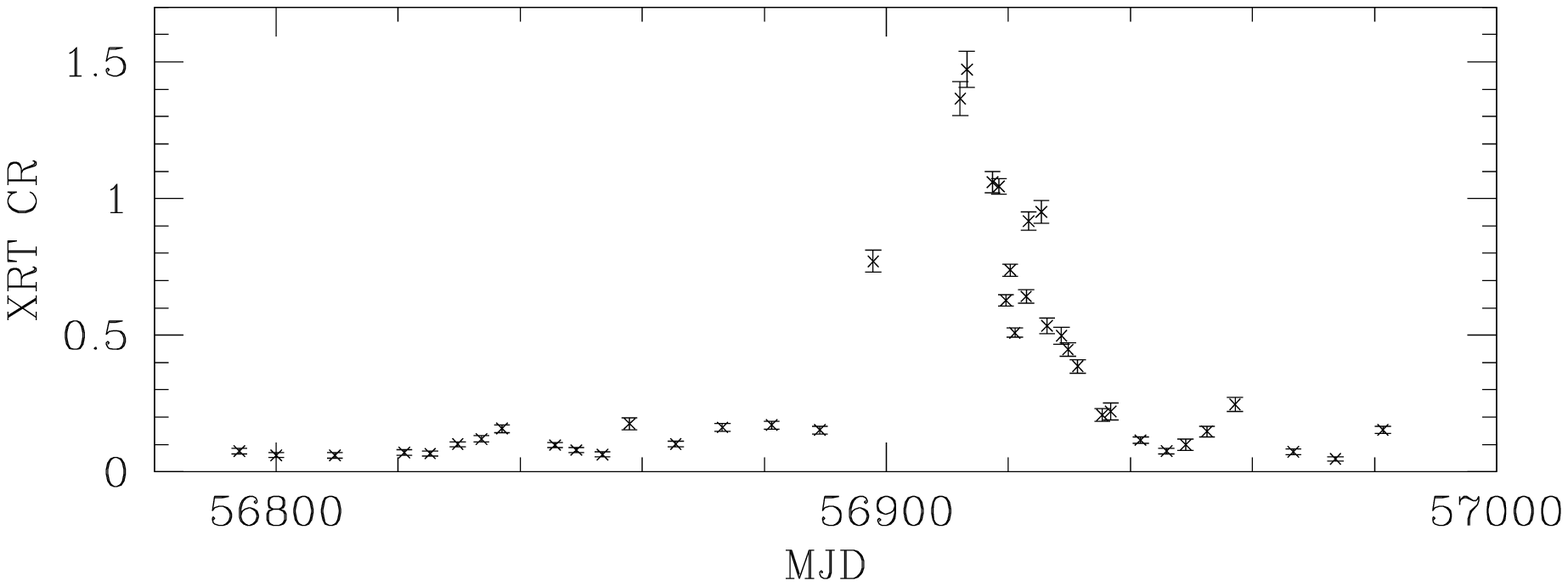}
\vspace*{-4.5 cm}
\caption{Zoom on the recent flaring activity of Mrk 335 which peaked in September 2014.}
\end{figure*}

\section{Mrk 335: the AGN that cannot make up its mind}

Mrk 335 is a prominent, nearby, bright Seyfert galaxy, which has been observed with
every major previous X-ray mission. \textit{Swift} discovered
a historic minimum flux state (Grupe et al. 2007). The long-term X-ray lightcurve
of Mrk 335 is shown in Fig. 1. While it initially spent most of the time in a higher state,
in recent years, it has increasingly often been found in X-ray weak states plus occasional flaring.
The 2014 lightcurve of Mrk 335 (Fig. 2) shows a strong flare, which peaked in Sept. 2014
followed by a rapid decline.  The peak flux reached a value not seen since 2008.
%

The intermediate and low states are characterized by strong spectral complexity.
These states were followed-up with deep \textit{XMM-Newton} and broad-band \textit{Suzaku}
spectroscopy. The \textit{XMM-Newton} observations
have revealed the presence of highly ionized absorption (Grupe et al. 2012, Longinotti
et al. 2013).
Broad-band \textit{Suzaku} spectroscopy (Gallo et al. 2015) of a recent
deep low state of Mrk 335 has shown that it is well
explained with a blurred reflection scenario (then requiring high black hole spin), or alternatively
with a partial-covering absorber (nearly Compton-thick in low state).

\begin{figure*}[th]
\centering
\includegraphics[width=3.5in]{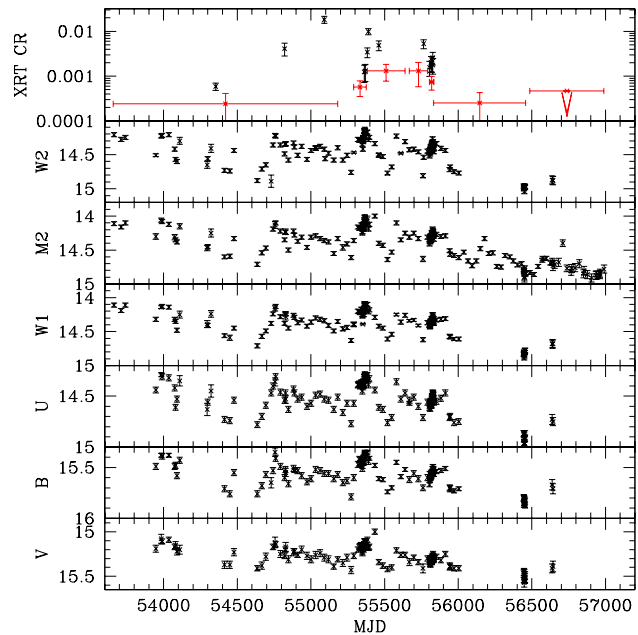}
\caption{\textit{Swift} long-term lightcurve of WPVS\,007 until December 2014. No more detections in
X-rays have occurred during the last year, and the UV shows an overall downward trend, too. }
\end{figure*}

\section{WPVS\,007: the AGN that disappeared}

WPVS\,007 almost vanished from the X-ray sky $\sim$25 years ago (Grupe et al. 1995b),
while its optical spectrum remains that of a narrow-line Seyfert 1 (NLS1) galaxy, and
unchanged. UV
spectroscopy (Leighly et al. 2009) revealed the likely cause of the X-ray faintness: the
onset and development of a strong broad-absorption line (BAL) flow,
remarkable for this low-mass galaxy.
\textit{Swift} monitoring (Grupe et al. 2013) discovered occasional flickering into X-ray
bright states, presumably due to leaky absorption, and high-amplitude UV variability
(Fig. 3).
The recovery into its ``normal state'' is yet to come, and will provide us with an excellent
opportunity of studying BAL processes on short time scales.

\section{IC\,3599: the AGN that re-appeared}


\noindent {\bf{\textit{Swift} discovery of a new outburst.}} The Seyfert galaxy IC\,3599
(Zwicky 159.034)  underwent a dramatic X-ray outburst during the ROSAT all-sky
survey,  accompanied by a strong brightening of the optical emission lines
which then faded again (Brandt et al. 1995, Grupe et al. 1995a,
Komossa \& Bade 1999). Several outburst scenarios were discussed, including
high-amplitude NLS1 variability, a disk instability, or a stellar tidal disruption
event (TDE). IC\,3599 was already known
as a Seyfert galaxy before the outburst (based on its narrow emission lines like [OIII]),
so it was difficult to make a strong case for a TDE,
because of the long-term presence of an accretion disk in that galaxy, and therefore the possibility
to link the high-amplitude variability to accretion processes (similar to, but more extreme than,
other AGN).

IC\,3599 had remained very faint ever since, but was recently discovered in
another outburst with \textit{Swift} (Fig. 4). Inspecting data
from the \textit{Catalina} sky survey (Drake et al. 2009),
we find that the
X-ray emission was accompanied by enhanced optical emission, which was already at similar brightness
2 years earlier (Fig. 4).
After we noticed the new high state, we triggered new observations
with \textit{Swift}, which show that the 2010 outburst is over,
and X-ray emission levels are down again
by a factor $\sim$100.  How does this second outburst constrain emission scenarios discussed earlier?
Which new ones do emerge? Below, we briefly raise some possibilities. Aspects of these models,
and the full data sets, are further discussed by Grupe et al. (2015, to be subm.).

\begin{figure}[b]
 \includegraphics[width=2.6in]{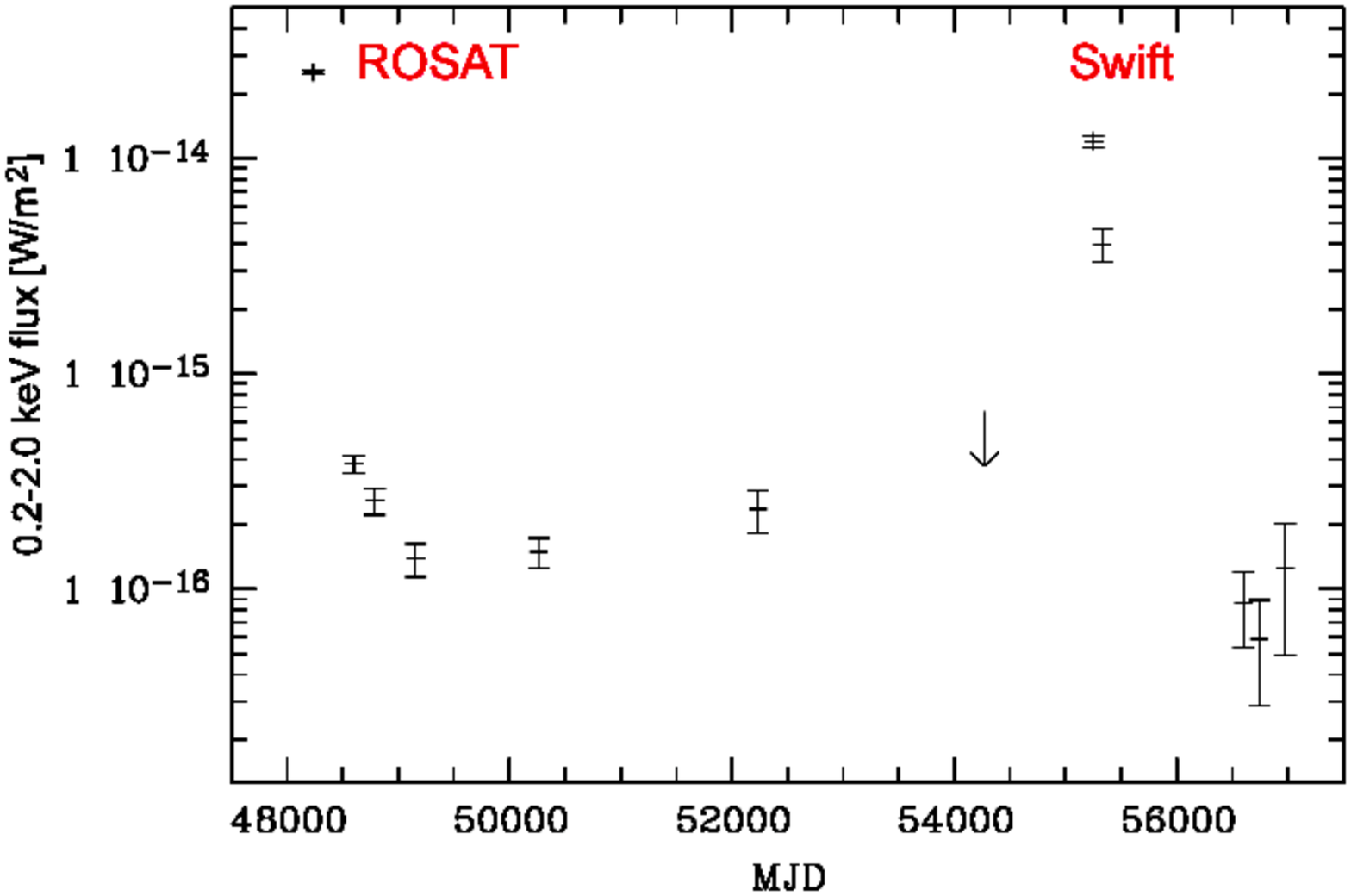}
\includegraphics[width=2.7in]{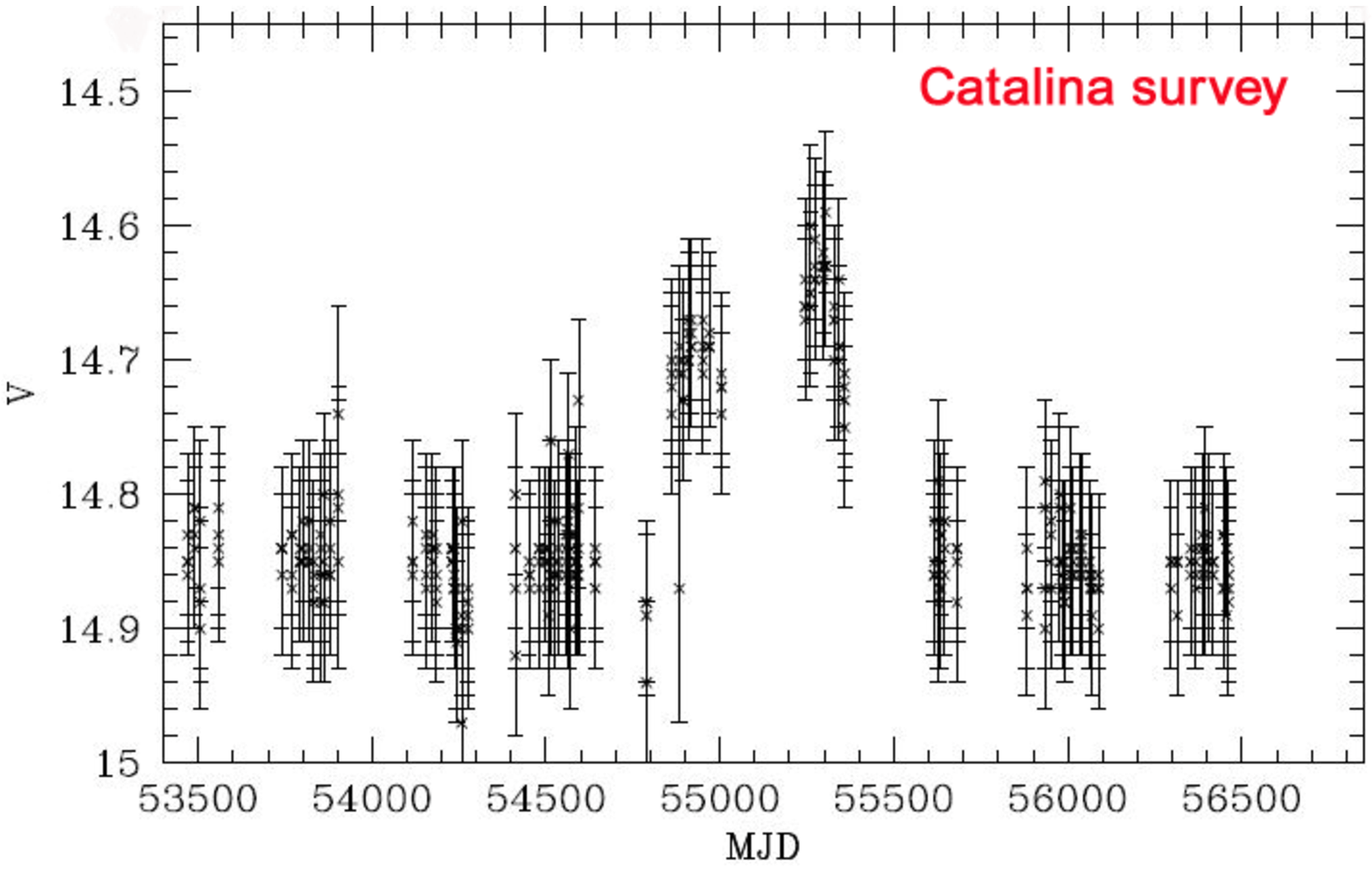}
 \caption{\textit{Left:} X-ray lightcurve of IC\,3599, starting with the initial
outburst detected with ROSAT. The last data points are from \textit{Swift}, which found
IC\,3599 back in a high state in 2010. We therefore triggered follow-ups with \textit{Swift},
which showed that flux levels have strongly decreased again.
\textit{Right:}
\textit{Catalina} survey lightcurve of IC\,3599, showing that the new X-ray peak was accompanied by enhanced
emission in the optical, which was already bright in 2008, and still (or again) in 2010. Increased
optical and UV emission was also seen with \textit{Swift} in 2010.
 }
\end{figure}


\noindent {\bf{TDE-related outburst scenarios.}} At first glance, the detection of
repeated flaring is unexpected, if the TDE interpretation was
correct. However, we note that recurrent outbursts or high states may also occur in TDE scenarios,
for instance when one of the following conditions is met: (1) If IC\,3599 hosts
a binary black hole or recoiling black hole, tidal disruption rates are temporarily
strongly boosted (e.g., Chen et al. 2009, Komossa \& Merritt 2008, 
Stone \& Loeb 2011), and a new disruption
event can occur within a decade. (2) If a TDE in IC\,3599 happened in a  binary SMBH,
its lightcurve would show characteristic
recurrent dips, since the presence of the secondary temporarily
interrupts the accretion stream on the primary
(Liu et al. 2009); a model successfully applied to the TDE lightcurve from SDSSJ120136.02+300305.5
(Liu et al. 2014). (3) Some very loosely bound clumps of stellar debris may return late to the BH causing
occasional additional accretion flares, even though it is unlikely that they produce a peak luminosity
similar to the earlier one. (4) Another variant of explaining the new lightcurve of IC\,3599 with a TDE
was presented by Mainetti et al. (2015) at this \textit{Swift} meeting.


\noindent {\bf{AGN-related outburst scenarios.}} Given that IC\,3599 hosts a long-lived
AGN, extreme processes in its accretion disk,
possibly due to instabilities, are a possible
cause of its recurrent outbursts. We note in passing that an extreme absorption event like the
one in WPVS\,007 is a very unlikely
explanation for the lightcurve of IC\,3599, because of the significant brightening of several optical
emission lines of IC\,3599, strongly arguing for a true outburst at that time.
If the outbursts continue repeating, the behaviour is reminiscent of OJ287 (e.g., Sillanp\"a\"a et al.
1988, Valtonen et al. 2008), and a binary
SMBH  might be responsible (i.e., a secondary BH interacting with the accretion disk around the primary while
orbiting).
Another mechanism which produces repeat outbursts is episodic stream-feeding of one of the
black holes in a binary SMBH system (Tanaka 2013) which recurs every orbit.
Ongoing monitoring with \textit{Swift} will provide tight new constraints on outburst models.

\section{XMM\,J061927.1$-$655311: giant AGN outburst or a TDE?}

XMM\,J0619$-$65 was caught in a high-amplitude flaring state (factor $>$140) with \textit{XMM-Newton}
($L_{\rm x,peak} \sim 10^{44}$ erg/s). Optical follow-up spectroscopy revealed low-level
AGN activity. The X-rays and UV, observed with \textit{Swift}, faded
subsequently (Saxton et al. 2014).
Most likely, we have seen an extreme case of AGN variability (a change in accretion rate).
Alternatively, a TDE occurred at the core of this galaxy.

\section{RXJ2314.9+2243: an extreme and radio-loud NLS1 galaxy}

RXJ2314.9+2243 is one of the few radio-loud NLS1 galaxies (Komossa et al. 2006), with
a tentative $\gamma$-ray detection (L. Foschini, priv.com.; Berton et al. 2015, in prep.).
With \textit{Swift}, we have measured its UV--X-ray spectral energy distribution
(SED) quasi-simultaneously for the first time
(Komossa et al. 2015). Its SED shows a broad hump from the IR to UV with a steep decline
in the UV, unlike radio-quiet NLS1 galaxies (Grupe et al. 2010), 
but consistent with a Synchrotron
origin of the emission.
Overall,
RXJ2314.9+2243 shares the dual properties of blazars and NLS1s exhibited by the few other
known radio-loud $\gamma$-ray-emitting NLS1 galaxies. Further, it shows a remarkably
strong outflow component in its broad [OIII] emission line
($v=1260$ km/s), and likely represents an extreme case of AGN-induced feedback
in the local universe (Komossa et al. 2015).



\end{document}